\documentclass[prb,aps,twocolumn,floatfix,showpacs]{revtex4}%
\usepackage{amssymb}
\usepackage{amsmath}
\usepackage{graphicx,color}
\usepackage{amssymb}
\usepackage{amsfonts}%
  \usepackage{color}
  \definecolor{darkgreen}{rgb}{0,0.5,0}
  \definecolor{green2}{rgb}{0,0.8,0}
  \definecolor{darkblue}{rgb}{0,0,0.5}
  \definecolor{darkred}{rgb}{0.7,0,0}
  \definecolor{darkyellow}{rgb}{0.7,0.7,0}

  \RequirePackage[hyperindex,colorlinks,bookmarksnumbered,plainpages=true]{hyperref}
  \hypersetup{linkcolor=blue,urlcolor=blue,citecolor=green2}
  \usepackage{hyperref}

\begin{document}

\title[ShortTitle]{Tunability of qubit Coulomb interaction: \\ Numerical analysis
     of top gate depletion in two-dimensional electron systems}
\author{A. Weichselbaum}
\author{S.~E. Ulloa}
\affiliation{ Department of Physics and Astronomy, Nanoscale and
Quantum Phenomena Institute, Ohio University, Athens, Ohio 45701}

\begin{abstract}
We investigate the tunability of electrostatic coupling between solid state
quantum dots as building blocks for quantum bits. Specifically, our analysis
is based upon two-dimensional electron systems (2DEG) and depletion by top
gates. We are interested in whether the Coulomb interaction between qubits can
be tuned by electrical means using screening effects. The systems under
investigation are analyzed numerically solving the Poisson equation in 3D via
relaxation techniques with optimized algorithms for an extended set of
boundary conditions. These include an open outer boundary, simulation of 2DEG
systems and dielectric boundaries like the surface of a physical sample. The
results show that for currently lithographically available feature sizes, the
Coulomb interaction between the quantum bits is weak in general due to
efficient screening in the planar geometry of 2DEG and top gates. The
evaluated values are on the order of $1 \,\mathrm{\mu eV}$. Moreover, while it
is not possible to turn off the qubit interaction completely, an effective
tunability on the order of $50\%$ is clearly realizable while maintaining an
intact quantum bit structure.

\end{abstract}
\volumeyear{year}
\volumenumber{number}
\issuenumber{number}
\eid{identifier}
\date[Date: ]{June 26, 2006}



\maketitle


\section{Introduction}

Quantum dot structures based upon localized regions of charge play an
important role in investigating quantum mechanical effects at the nanoscale.
\cite{Fujisawa03,Fuhrer03,Pioda04} Despite the short decoherence times,
considerable progress is made in utilizing the charge degree of freedom for
coherent quantum processes with possible application in quantum information.
\cite{Fujisawa04,Toth01,Jefferson96} In particular, electrostatically-defined
quantum dots based on depletion of two-dimensional electron systems (2DEG), play an
important role due to their variability, as the charge ``drop'' in the dot can
be controlled in shape and size by applying well-controlled gate voltages.

In this paper we consider a linear array of double dots as shown in
Fig.~\ref{fig0}. With different arrangement of voltage gates,
quantum dots are formed and arranged in pairs to define charge
qubits. Every pair is defined with a tunnel contact (quantum point
contact, QPC) between the two dots, which can be tuned by an
external gate voltage. With the QPC in a weakly transparent regime,
the charges are well localized in their respective quantum dot, and
the dot occupation is characterized by a well-defined number of
electrons. The system can then be tuned to the energetically
degenerate case, where an extra electron can reside on either
quantum dot symmetrically. The dynamics of such a system is
well-described by the two qubit states $\left\vert L\right\rangle $
and $\left\vert R\right\rangle $, with the extra electron either on
the L(eft) or on the R(ight) dot. These two states form the natural
basis to define the two charge qubit configurations easily
controlled by the gates that define the quantum dots and the QPC.
\cite{Gorman05}

\begin{figure}[ptb]
\begin{center}
\end{center}
\caption{Linear array of double quantum dots.  Charge is exchanged
  between the dots in each pair but different pairs are only coupled
  electrostatically.  Defining gates are not shown.}%
\label{fig0}%
\end{figure}

As different dot pairs are used to define charge qubits via local
gate voltages, their states are well defined and controlled
experimentally.  However, it is crucial to be able to control the
interactions between neighboring qubits if one is to implement the
quantum gate operations required in quantum computation.
\cite{Nielsen00} We are thus interested here in the energetics of
different charge configurations in the double-dot array and the {\em
tunability} of the coupling between charge qubits via purely electrostatic
means.  Notice that including the interaction with neighboring
qubits results in an effective dipole-dipole interaction (the common
$ZZ$ interaction in solid state qubits in a spin notation).  This
causes the states in the qubit to align according to their
neighbors, and misalignments to be energetically costly.  We
investigate the energetics of different qubit configurations and how
one can stabilize different states by purely electrostatic means.

For typical nanoscopic devices with many (or at least a few)
electrons in each of the electrostatically-defined regions, the
charge distribution and the major energy scales are described to a
good approximation by classical electrostatics. \cite{Marchi04} Due
to the strong electric fields generated by segregating charge in a
2DEG, the Coulomb energy is the dominant energy scale. Thus, it is
important to know the electrostatics of the system if one expects a
good quantitative description thereof. We present here a detailed
study of the electrostatic coupling in the quantum dot array of
Fig.~\ref{fig0}. We find that one can substantially modify the qubit
couplings by proper gate geometry and voltage control ($\approx
50\%$).  However, we also find that it is not possible to completely
turn off the coupling among dots, as one would need in order to
allow for total decoupling of the system. Our studies utilize an
efficient three-dimensional electrostatic code that uses
higher-order grid relaxation and fast Fourier transform algorithms.
This allows for accurate and realistic modelling of the quantum dot
array of interest here, as well as a variety of other structures.
Effects such as depletion in a 2DEG due to shallow etch
\cite{Vancura03} or deep etch \cite{Klein04} have been investigated
with our code and the results agree well with experimental data and
other simulations built on Schr\"{o}dinger Poisson solvers.
\cite{Snider90}

As we mentioned above, we are interested in the tunability of the
coupling between charge qubits. Electrostatically, tunability is directly
related to changes in the geometry of the system, i.e. the regions
were there is charge, since a constant geometry of electrically
isolated regions implies a constant capacitance matrix and constant
electrostatic coupling.  Application of gate voltages produces
changes in the lateral charge depletion on the 2DEG system,
drastically changing the electric field line distribution around the
area of interest. In the quantum dot structures analyzed here, this
includes field lines that emerge out of the dots through the
surface, as well as throughout the semiconducting material including
the depleted 2DEG region.  Notice, however, that depletion of charge
in the 2DEG already suggests that the local electric field is strong
with dominant field lines vertical to the 2DEG layer. Small local
changes of charge or rearrangements are likely \textit{not} to
affect the interdot coupling strongly. This conclusion is not
trivially reached, due to the complexity of the dot array and gate
geometry, which includes subtle screening effects. Still, this
intuitive picture turns out to be correct, as will be demonstrated
in detail in the following.

\section{Simulation of 2DEG}

2DEG systems are ``volatile'' by design, i.e. the electric charge
can be depleted or accumulated by external means using appropriately
fabricated conducting leads acting as voltage gates. Thus the 2DEG
regions containing electric charge density change dynamically with
the voltage patterns applied to the gates. For a flexible and
reliable numerical analysis it is thus necessary that the 2DEG
boundary is treated dynamically. The plane of the 2DEG may contain
several independent regions of charge that can be addressed
individually. Moreover, well isolated regions of charge have a
floating potential, and this situation remains even if weak tunnel
junctions are present because of charge quantization. These
considerations introduce a set of specific boundary conditions
related to 2DEG systems. They are readily realized in our numerical
code and explained in detail in [\onlinecite{Wb2004}]. Moreover, as
from a numerical point of view we are dealing with a finite system,
the potential on the outer boundary floats freely and the system
must be solved self-consistently. For more detail in that respect
the reader is also referred to the Appendix.

The primary boundary condition for the 2DEG as it turns out is
surprisingly simple local update prescription built in easily with
the relaxation sweeps: \cite{Wb2004} when coming across a 2DEG grid
point, say point $i$, with the 2DEG layer kept from the exterior at
the potential $V_{\mathrm{2DEG}}$, then proceed as follows - first,
update the potential $V_{i}$ locally as in homogenous space. Second,
if $V_{i}^{\mathrm{new}}>V_{\mathrm{2DEG}}$, take
$V_{i}=V_{\mathrm{2DEG}}$ otherwise accept the calculated
$V_{i}^{\mathrm{new}}$. The result is that the regions where $V_{i}
=V_{\mathrm{2DEG}}$ contain electronic charge while the regions in
the 2DEG with $V_{i}<V_{\mathrm{2DEG}}$ are depleted. The basic idea
underlying this prescription is simply that for the electronic
charges $q_{i}<0$ in the 2DEG it is energetically favorable to go to
a potential that is higher then $V_{\mathrm{2DEG}}$. The effect is
that charges gather in these regions until the potential
equilibrates at $V_{\mathrm{2DEG}}$.

\begin{figure}[ptb]
\begin{center}
\end{center}
\caption{(Electric potential $V(z)$ in a parallel plate geometry to
demonstrate depletion in 2DEG -- the material below the surface (to
the right of the $\sigma_1$ layer) is considered uniform with a
dielectric constant of $\varepsilon$. The charge densities on the
three planes shown are $\sigma_{1}$, $\sigma_{p}$ (p for the
\textit{ positive} charge of the donor layer) and $\sigma_{2}$,
respectively. $V_{0}$ is the total potential difference across the
whole stack related to the pinning of the Fermi level at the
surface.}
\label{fig_depletion}%
\end{figure}

The gross features of the charge distribution in a realistic sample
with a uniform 2DEG system buried in close vicinity to the surface
can be well understood by a parallel plate capacitor arrangement as shown
in Fig.~\ref{fig_depletion}. $\sigma_{1}$ is the charge density at
the free surface. A distance $d_{1}$ below lies the donor layer
(``delta doping'') with a doping density $\sigma_{p}\equiv en_{D}$,
followed by the 2DEG separated by a distance $d_{2}$ from the
donors. The variables $d_{1}$ and $d_{2}$ are parameters that can be
adjusted to account for the finite width of each of the three
layers. The electrical voltage $V_{0}$ of the surface with respect
to the 2DEG is fixed as a consequence of the pinning of the Fermi
energy at the surface and makes up for the difference in chemical
potential due to surface states. For the case of Ga(Al)As, for
example, the Fermi level is pinned to mid-gap,\cite{Kawaharazuka02}
and so the potential $V_{0}$ is chosen to be fixed at
$V_{0}\simeq-0.7\,\mathrm{V}$ (note that the 2DEG is considered
grounded).

The charges at the surface and in the 2DEG rearrange and equilibrate when the
temperature is high enough, typically at room temperature. This leads to the
energetically most favorable charge distribution that shares the charges
available from the donor layer between the surface and the 2DEG such, that the
electrical field above the surface and below the 2DEG of the sample is zero.
The remaining adjustable parameter in a real system is then the
distance $d_{1}$ which can be reduced by etching. The critical
distance $d_{1}^{\ast}$ for depletion of the 2DEG layer below is given
by $d_{1}^{crit}=-V_{0}\varepsilon\varepsilon_{0}/en_{D}$ which is independent
of the distance $d_{2}$ (note that $V_{0}$ is negative here).

With clearly separated layers of uniform charge, this capacitor model can be
expected to provide a good description of charge distribution. The quantum
mechanical effects are implicitly present in the effective parameters $d_{1}$
and $d_{2}$ as well as in effects such as the pinning of the Fermi energy at the
surface.  To this extent, this semiclassical description of the system
provides an accurate picture. \cite{Marchi04}

\section{Energetics of double dot qubits \label{sec.doubledots}}

Double dot systems are created in the 2DEG via depletion through
negatively biased metallic top gates on the surface of the
structure, e.g. fabricated by lithographic means. The typical system
analyzed here is shown together with its numerical results in
Fig.~\ref{fig_G12_quV}. The relatively large number of gates
defining the structure keeps the system flexible. However, the exact
number of gates and their respective arrangement are arbitrary to a
certain extent and may be simplified by fewer gates in a specific
experimental setup.

Every double dot system (qubit) in our design has altogether six
gates defining the dots, including a plunger gate for every dot, and
two gates that tune the tunnel barrier (QPC) between the two dots
defining the qubit. One of the two QPC gates for a pair of dots may
eventually be merged with one of the gates confining the dot.

The set of top gates chosen can be seen in panel (a) of
Fig.~\ref{fig_G12_quV}. The charge distribution shown with biased
gates clearly outlines their geometry. The red shading (negative
charge density) results from the negative bias voltages applied.
Thus the more negative the bias, the larger the accumulated negative
charge on that gate.

The potentials on the top gates were chosen as follows depending on
the local degree of depletion that was required: $-2.5\,\mathrm{V}$
on the plunger gates for every dot, $-5.2\,\mathrm{V}$ for the QPC
between a pair of dots, $-2.4\,\mathrm{V}$ for the transition region
between pairs of quantum dots, except for the center transition
whose gate potential was alternatively also set to $0\,\mathrm{V}$.
For the other confining potentials a value of $-4.5\,\mathrm{V}$ was
taken. The remaining (geometrical) parameters are: pinning of the
surface potential $V_{0}=-0.75\,\mathrm{V}$, dielectric constant
$\epsilon=12$, grid spacing $h=12\,\mathrm{nm}$,
$d_{1}=d_{2}=36\,\mathrm{nm}$, and
$n_{D}=4\cdot10^{16}\,\mathrm{m^{\text{-2}}}$, resulting in a charge
distribution ratio between the surface and the 2DEG of about $2:1$.

\begin{figure}[pt]
\begin{center}
\end{center}
\caption{(Color online) Simulation of 2DEG with top gates and
applied gate voltages. The center gates (labeled $V_{c}$ in panel a)
extend in between the qubit pair such that the 2DEG underneath the
gate $V_{c}$ can be depleted at will (see panels b and c). The color
coding for charge distributions is consistently taken such that red
corresponds to negative and blue to positive charge density (in gray
scale all shades are to be considered red except for the dark gray
regions in panel d). a) Outline of top gates by showing the charge
distribution on these gates. The $1\,\mathrm{\mu m}$ bar shows
chosen length scale. b) Charge distribution in the 2DEG with the
center gate at the same voltage as the gates separating the other
qubits. c) Same as (b) but with the voltage on the center gate
lifted such that additional charge accumulation is allowed in the
2DEG underneath the gate. d) Differences in the charge distribution
for the setup in (c) resulting from small variations on the plunger
gate voltages chosen such that the total change of charge within one
dot is $\pm\,e/2$. e) Free electrostatic energy of the different
charge configurations shown in (d) taken
relative to the first configuration $\#1$.}%
\label{fig_G12_quV}%
\end{figure}

In the setup of Fig.~\ref{fig_G12_quV}, the quantum dots contain
about $200$ electrons. In the following, the two adjacent qubits in
the center of Fig.~\ref{fig_G12_quV} are considered in more detail,
and the outer qubits allow us to examine the effect of qubit pair
interaction. The plunger gate for individual dots was fine tuned
self-consistently at the end of the initialization process to
produce half-integer occupancy on the dot. With this it is then an
easy numerical task to achieve equivalent integer charge
distributions for a single qubit with one extra electron
symmetrically on either of the dots forming the qubit labelled as
the two qubit states $\left\vert L\right\rangle $ and $\left\vert
R\right\rangle $. More specifically, with the gate voltages frozen
to their initial configuration for half-integer occupation of the
dots, the charge configuration can now be altered to the possible
different integer value charge configurations, shifting half of an
electron in a quantum dot to the other or vice versa.

As indicated in Fig.~\ref{fig_G12_quV}, the gate in the center is
introduced for the purpose of screening the qubit-qubit interaction
through depletion or accumulation of charge in between the qubits,
thus introducing the change of geometry of the charged regions
necessary for tunability of the qubit interaction. The two resulting
configurations analyzed are shown in panels (b+c) for the gate
voltages $V_{c}=-2.4\,\mathrm{V}$ and $V_{c}=0\,\mathrm{V} $,
respectively. For the setup in panel (b), the different qubit states
resulting from single electron hopping on the two qubits in the
center are presented in the panels (d). For better visual contrast,
the difference in the charge configuration $\Delta q\left(
\vec{r}\right)  $ in the layer of the 2DEG with respect to the
initial half-integer setup is shown. When summing up $\Delta q\left(
\vec{r}\right)  $ over a single dot, then the total difference in
charge $\Delta Q_{i}$ for dot $i$ is $\left\vert \Delta
Q_{i}\right\vert =e/2$ up to numerical precision. Note that the
influence on the outer qubits, i.e. the qudot pair on the very left
and the very right as seen in panels (b+c) is minimal, and no
appreciable charge rearrangement occurs there due to the charge
alterations in the two middle qubits.

From the numerical evaluation of the electrostatics, we obtain the
potential of the quantum dot array together with the total charge on
every dot. For a calculated set of inital values $q_{a}$ and $V_{a}$
for the charge and the potential on the individual dots, the
electrostatic free energy is evaluated using $\Delta
W_{ab}=\int\limits_{q_{a}}^{q_{b}}V\left(  q^{\prime};V_{g}\right)
dq^{\prime}$. The vector $V_{g}$ stands for a set of external
potentials held at constant voltage. For the case of weak changes in
the systems electrical geometry, the total change in free energy for
different charge configurations is given by \cite{Wb2004}
\begin{equation}
\Delta W_{ab}\simeq\Delta q~\frac{1}{2}\left(  V_{a}+V_{b}\right)  \text{,}
\label{e3_free_energy}%
\end{equation}
where $\Delta q\equiv q_{b}-q_{a}$. For a constant geometry
(capacitance matrix) this equation holds exactly. However, in the
relevant cases here where gate voltages do change the geometry,
namely the shape of the quantum dots as well as the remaining
extended regions of charge in the 2DEG, Eq.~(\ref{e3_free_energy})
only holds to first order in small variations of the geometry. Yet
this is definitely the case for the single-electron hopping
processes considered here.

The qubit-qubit interaction energy for the configurations in panel
(d) is then evaluated by means of Eq.~(\ref{e3_free_energy}) with
the results shown in panel (e) of Fig.~\ref{fig_G12_quV}. The
quantum dots have been intentionally designed not to be exactly the
same, resulting in small relative charge variations (maximum
deviation of two electrons between the dots). This variation is also
reflected in the slight assymetry seen in panel (e) between the
ideally equivalent cases shown in the panels (d). For the setup in
panel (b) with the charge depleted under the center gate, the
qubit-qubit interaction energy is given by $\Delta
E\simeq1.17\,\mathrm{\mu eV}$. By lifting the potential on the
center gate from $V_{c}=-2.4\,\mathrm{V}$ to $V_{c}=0\,\mathrm{V}$,
thus allowing extra charge to screen the qubit-qubit interaction
(see middle of panel c), the interaction energy is reduced to
$\Delta E\simeq0.66\,\mathrm{\mu eV}$ which is about half the former
value.

The quantum dot spacing is about $300\,\mathrm{nm}$ from center to
center of charge within a qubit, and $430\,\mathrm{nm}$ between
neighboring qubits. Considering the initial half-integer
configuration as neutral, i.e. the negatively charged qudots as
screened by the environment, then a simple classical estimate of the
energy scale for the qubit-qubit interaction by shifting half of an
electron charge in between the qubit gives $65\,\mathrm{\mu eV}$,
which is two orders of magnitude larger than the actual realization
of dots formed by depletion. This strong suppression of the
interqubit interaction is clearly due to the large metallic top
gates which efficiently screen the slightly rearranged charge
configurations closeby.

A similar calculation was performed with the center gate stretching only
half-way through the qubit-qubit intermediate regions. The results are shown in
Fig.~\ref{fig_S14_quV}. The average distance between the qudots is slightly
different from the previous case ($320\,\mathrm{nm}$ intraqubit vs.\
$430\,\mathrm{nm}$ interqubit separation). The confining potentials
on the top gates are slightly weaker, allowing larger dots with about $300$
electrons each. The remaining parameters are the same as in the
previous case.

The dots are completely symmetric in this case, reflected in the
energy of the equivalent charge configurations in panel (d) plotted
in panel (e). Again, the qubit-qubit interaction energy for each
configuration in panel (d) was evaluated using
Eq.~(\ref{e3_free_energy}), and results shown in panel
\ref{fig_G12_quV}(e). For the setup in panel (b) with the charge
depleted all along the center gate, the qubit-qubit interaction
energy is given by $\Delta E\simeq0.71\,\mathrm{\mu eV}$. By raising
the potential on the center gate from $V_{c}=-2.5\,\mathrm{V}$ to
$V_{c}=-0.75\,\mathrm{V}$, the interaction energy is reduced to
$\Delta E\simeq0.29\,\mathrm{\mu eV}$ which is less than half the
former value, but still clearly present. The simple classical
estimate similar to the previous example with the extra charge in
the qudots positioned around the center of charge leads to
$71\,\mathrm{\mu eV}$, again two orders of magnitude larger than the
energies in the actual realization.

\begin{figure}[pb]
\begin{center}
\end{center}
\caption{(Color online) Same as Fig.~\ref{fig_G12_quV} but with the
center gate (labeled $V_{c}$ in panel a) extending only halfway
through from top to bottom. This structure has also slightly
increased quantum dot spacing and size from that in
Fig.~\ref{fig_G12_quV} . The color coding for charge distributions
is the same, with red for negative and blue for positive charge
density (in gray scale all shades are to be considered red except
for the dark gray regions in panel d). See description of panels
(a-e) in caption of Fig.~\ref{fig_G12_quV}.}
\label{fig_S14_quV}%
\end{figure}

\section{Conclusions}

The tunability of qubit interaction based on electrostatic gate
screening has been analyzed numerically. Our program uses an
efficient relaxation algorithm, and efficiently describes general 3D
structures, including the ability to handle an extended set of
boundary conditions, specifically appropriate for the treatment of
2DEG systems and including dielectric effects.

The typical systems analyzed are based on 2DEG geometries derived
from GaAs/GaAlAs structures and depletion via top gates. For a chain
of qudot pairs -- the qubits -- the region in the 2DEG between
qubits is depleted at will, allowing for tunable screening and
control of the nearest neighbor qubit interaction. The tunability
derived for realistic gate voltage changes is on the order of
$50\%$. The interaction energy can thus be clearly modulated.
However, we find that it cannot be turned {\em off} for moderate
voltages, as it would be desirable for the on/off-switching of the
interaction of ideal qubits. We believe that this general trend does
not only hold for the few systems analyzed in this paper. The
tunability based on charge rearrangement and screening of Coulomb
interaction on charged-localized quantum dots is definitely
possible, but limited in efficiency.  It may indeed be necessary to
implement dynamic pulse control techniques, such as the
``bang-bang'' approach or others discussed for qubits recently.
\cite{Tian00}

\begin{acknowledgements}
We acknowledge helpful discussions with T. Ihn and the whole group of
K.~Ensslin, as well as their hospitality.  The work was partially
supported by NSF Grant NIRT 0103034, and the Condensed
Matter and Surface Sciences Program at Ohio University.
\end{acknowledgements}

\appendix
\section*{Appendix -- Relaxation on the grid and open boundaries}

The algorithm used in this paper is based on a higher-order
relaxation algorithm for a finite physical system as described by
the authors in an earlier paper. \cite{Wb03} Major modifications are
significantly optimized update of the outer boundary, and an
extended set of additional boundary conditions mainly related to
2DEG systems. As later ones represent dielectric media it is also
clear that a proper treatment of surfaces requires a proper
incorporation of the dielectric constant. A detailed description can
be found in reference [\onlinecite{Wb2004}].

Solving the electrostatics in 3D for a finite system with an
\textit{open} outer boundary may not allow for a simple boundary
condition such as constant potential on the outer surface. With
uniform grid spacing, one is bound to a finite total grid size,
which in 3D is even more stringent. Therefore the region under
investigation cannot assume constant potential on the outer
boundary. However, by having direct access to the local charge
distribution throughout the grid during relaxation, one can
\emph{calculate} the potential on the outer boundary. The naive way
of summing up charge over distance $V_{i}\sim\sum _{j}q_{i}/d_{ij}$,
however, is computationally expensive, to the extent that the actual
relaxation process becomes much faster compared than the
self-consistent update of the outer boundary.

The reason for this lies in the folding of the charge versus the
relative distance. The structure of the sum suggests the use of fast
Fourier transform (FFT) algorithms in 3D \cite{Korsmeyer98} where
the folding is replaced by a single product for each Fourier
component. The fact that FFT automatically implies periodic boundary
conditions which may introduce artificial effects can be
circumvented by doubling the system size in every dimension.
\cite{Hockney88} Despite the largely enhanced number of grid points
in 3D, the resulting update of the outer boundary is drastically two
orders of magnitude faster than the original way of carrying out the
sum in real space.

Typical materials such as Ga(Al)As or Al$_{2}$O$_{3}$ have a large
dielectric constant of order $\varepsilon=12$. This strongly affects
the screening of charges such as those in the 2DEG or surface
charges, and it is absolutely crucial to include the dielectric
properties in these calculations. The numerical implementation goes
along the lines of algorithms found in the literature,
\cite{Klapper86,Davis91} and it is discussed in detail in
[\onlinecite{Wb2004}]. The main idea is that the local relaxation of
the potential must be done under the constraint that there is no
free charge in bulk dielectric media. This leads to the
incorporation of the dielectric constant into the weights of the
potential average over neighboring sites,
$V_{i}^{\mathrm{new}}=\sum_{j\in \mathrm{NN}}
\epsilon_{j}q_{j}/\sum_{j\in \mathrm{NN}}\epsilon_{j}$. One should
be aware that the dielectric constant $\epsilon_{j}$ has to be
evaluated midway between the grid point $i$ under consideration and
the neighboring point $j$. \cite{Wb2004} Finally, note that the
subsequent calculation of the source $q_{i}$ from the potential
\textit{without }weighting it by the dielectric constant gives the
\textit{induced} charge at dielectric boundaries (note that in
uniform dielectric media, the dielectric constant drops out in the
averaging process for the local update of $V_{i}^{\mathrm{new}}$ and
so the induced charge equals zero by construction).

\bibliographystyle{amsplain}

\providecommand{\bysame}{\leavevmode\hbox
to3em{\hrulefill}\thinspace}
\providecommand{\MR}{\relax\ifhmode\unskip\space\fi MR }
\providecommand{\MRhref}[2]{%
  \href{http://www.ams.org/mathscinet-getitem?mr=#1}{#2}
} \providecommand{\href}[2]{#2}

\end{document}